\documentclass[pre,twocolumn,10pt,aps,longbibliography]{revtex4-1}
\usepackage{amssymb}
\usepackage{amsmath}

\usepackage{graphicx}
\usepackage{dcolumn}
\usepackage{bm}
\usepackage{subfig}

\usepackage{multirow}

\usepackage{color}

\begin{document}

\preprint{APS/123-QED}

\title{Effect of energy spectrum law on clustering patterns for
inertial particles subjected to gravity in Kinematic Simulation}
\author{F. C. G. A. Nicolleau}
\affiliation{Sheffield Fluid Mechanics Group, Department of Mechanical Engineering, The University of
Sheffield, Sheffield, United Kingdom}
\affiliation{Sorbonne Universit\'es, Universit\'e Pierre et Marie Curie, Paris 6 and Institut Jean le Rond d’Alembert, CNRS UMR 7190, Paris, France}
\email{corresponding author: F.Nicolleau@Sheffield.ac.uk}
\author{M. Farhan}
\affiliation{Department of Mechanical Engineering of the
University of Engineering and Technology Lahore, Pakistan}
\affiliation{Sheffield Fluid Mechanics Group, Department of Mechanical Engineering, The University of
Sheffield, Sheffield, United Kingdom}
\author{A. F. Nowakowski}
\affiliation{Sheffield Fluid Mechanics Group, Department of Mechanical Engineering, The University of
Sheffield, Sheffield, United Kingdom \mbox{ }}

\date{\today}


\begin{abstract}
We study the clustering of inertial particles using a periodic kinematic simulation.
Particles clustering is observed for different pairs of Stokes number and Froude number and
different spectral power laws ($1.4 \leqslant p \leqslant 2.1$).
The main focus is to identify and then quantify the effect of $p$ on the clustering attractor - by attractor we mean the set of points in the physical space where the particles settle when time tends to
infinity. It is observed that spectral power laws can have a dramatic effect on the attractor shape.
In particular, we observed a new attractor type which was not present in previous studies for Kolmogorov spectra ($p=5/3$).
\end{abstract}


\maketitle

\section{Introduction}

It is important to understand the particle clustering mechanisms in order to explore, identify and possibly monitor some natural
or hand-made mixing processes such as those causing rain formation \cite{Falkovich-al-2002,Falkovich-Pumir-2004,Woittiez-et-al-2008}, sediment
transportation \cite{Pan-et-al-2011}, fuel mixing and combustion.
The kind of turbulence found in any of these examples can be far from that underlying the observation of a classical Kolmogorov spectrum, as such examples can involve for
example stratification or rotation effects \cite{Matulka-et-al-2016}.

The effect of gravity cannot be neglected and was
analysed in previous studies where it has been shown to have a major impact on the shape and topology of the particle clusturing
\cite{Park-Lee-2014,Gustavsson-al-2014,Bec-et-al-2014,Farhan-et-al-2015}.
Spectral laws have also been shown to govern the particles separation
\cite{Morel-Larcheveque1974,Fung-Vassilicos-1998,Nicolleau-Yu-2004,Nicolleau-Nowakowski-2011}.
In the present paper we analyse the combination of both effects by extending our investigation of the effect of gravity to non-Kolmogorov energy
spectra.
\\[2ex]
This is of practical and theoretical interest.
Though the $-5/3$ spectrum turbulence has a general character and is often
observed in nature even when Kolmogorov assumptions are not
met (see e.g \cite{Laizet-et-al-2015}),
other sprectral laws are also observed in particular in two dimensional flows (e.g. plasmas and geophysical flows) where
a $-3$ spectral power law can be observed.
(See e.g. \cite{Sagaut-Cambon-2008} for an overview of non isotropic flows.)
From a theoretical point of view it is also interesting
to look at the effect of departing from the `classical'
$-5/3$ spectral law. For example in \cite{Nicolleau-Nowakowski-2011} it was observed that the $-5/3$ spectrum
corresponded to an extremum in the discrepancy between
theory and KS prediction.
\\[2ex]
Here following the work of \cite{Farhan-et-al-2015}, an initially uniformly distributed cloud of particles is tracked in a synthetic field mimicking turbulent flows (Kinematic Simulation). Clustering consists in the concentration of the cloud in some
regions of the physical space leading to a very inhomogeneous distribution of particles.
%
\\[2ex]
There are different ways to analyse particle clustering in turbulent flow and Direct Numerical Simulation
(DNS) is the most widely used method (e.g. \cite{Cencini-al-2006,Saw-et-al-2012,Falkovich-Pumir-2004}).
There are several reasons to use Kinematic Simulation for the study of particle
clustering \cite{Farhan-et-al-2015} but the main advantage for our study is that in this synthetic model the energy
distribution is an input variable. So there is no need for complex
forcing methods to create a particular turbulence energy distribution.
\\[2ex]
The clustering mechanism would be different in the inertial or dissipation range
of  turbulent flow \cite{Bec-2007}. In our paper we only study the effect of the scales in the inertial
range as this is possible by using a synthetic model where forcing and dissipation are not needed to develop
an inertial range.
\\[2ex]
Though there is no particular difficulty in considering particles with different inertia in Kinematic
Simulation, this study is limited to particles having the same inertia.
Furthermore, the particles are considered small enough so that they neither affect the flow  nor interact with
each other (one-way coupling).
\\[2ex]
The positions of particles are monitored as a function of time and a Lagrangian
attractor is observed for some cases. That is, the initially homogeneously distributed cloud of particles will end in a set
of loci that does not evolve any further. The particles move within that set which we call `Lagrangian attractor' and its dependence on $St$ and $Fr$ numbers is studied.

We only consider attractors with integer dimensions (one-dimensional and two-dimensional structures) which are easy to identify.
To compare our results with the reference case $p=5/3$ in \cite{Farhan-et-al-2015} we use the
nearest-neighbour distance analysis to identify the integer dimensions of the Lagrangian attractors
while varying the power law ($p$).
\\[2ex]
The paper is organised as follows: in \S~\ref{method} we introduce the KS model,\ its notations and its
parameters. The different kinds of Lagrangian attractor are discussed and introduced in \S~\ref{sectionIII}.
The effect of the spectral law is introduced in \S~\ref{sectionIV}.
A quantitative analysis is conducted in \S~\ref{sectionV}. Section~\ref{seconcl} summarises our main
conclusions.

\section{Kinematic simulation technique}
\label{method}

Kinematic Simulation (KS) is a particular case of synthetic turbulence where the focus is on particle's
trajectory at the expense of solving the Navier-Stokes equation. An analytical formula `synthetic flow' is
used for the Eulerian flow field.
Though the synthetic turbulence retains less information than the whole flow contains, its success relies on keeping what is paramount for the Lagrangian story.

The simplicity of the KS model excludes some features of real turbulent flows
but captures the part of the physics which is required to perform Lagrangian particle analysis.

KS modelling has been successfully employed and validated
\cite{Fung-al-1992,Elliott-Majda1996,Malik-Vassilicos1999}. This kind of simulation is much less
computing-time consuming than DNS, which is important for the present study where we need to run many cases
(more than a 1000 cases for 100 turnover times). Each case corresponds to a given $St$, $Fr$, $p$ and time and
involves 15625 particles.

With synthetic simulations, one can develop models where turbulence ingredients and complexity can be added
step by step helping to understand their respective importance. These synthetic models can be a useful
complement to Direct Numerical Simulation. In particular with KS it is possible to play with the spectral law
\cite{Nicolleau-Nowakowski-2011} and its consequences in terms of particle's dispersion.
\\[2ex]
As we are not interested in two-particle dispersion, we limit our study to the scale ratio ${k_i}_{max}/{k_i}_{min}=9$ \footnote{$i=1$, 2 or 3}
used in \cite{Farhan-et-al-2015}.
\\[2ex]
In KS, the computational task reduces to the calculation of each particle trajectory.
This trajectory is, for a given initial condition, $\mathbf{X}_0$, solution of the
differential equation set:
\begin{eqnarray}
{d\mathbf{X} \over dt} &=& \mathbf{V}(t)
\\
{d\mathbf{V} \over dt} &=& \mathfrak{F}(\mathbf{u}_{E}({\mathbf{X},t),\mathbf{V},t}) \label{a1}
\end{eqnarray}
where $\mathbf{X}(t)$ is the particle's position, $\mathbf{V}(t)$ its Lagrangian velocity and ${\bf u}_{E}$
the analytical Eulerian velocity used in KS. $\mathfrak{F}$ is a function relating the Lagrangian acceleration
to the Eulerian and Lagrangian velocities.

In KS ${\bf u}_{E}$ takes the form of a truncated Fourier series, sum of $N_k=N^3$ Fourier modes:
\begin{equation}
\mathbf{u}(\mathbf{x})=\sum_{i=1}^{N}\sum_{j=1}^{N}\sum_{l=1}^{N}
\mathbf{a}_{ijl} \cos (\mathbf{k}_{ijl} \cdot \mathbf{x})+
\mathbf{b}_{ijl} \sin (\mathbf{k}_{ijl} \cdot \mathbf{x})
\label{EqKSfield}
\end{equation}
where $\mathbf{a_{ijl}}$ and $\mathbf{b_{ijl}}$ are the decomposition coefficients corresponding to the
wavevector $\mathbf{k_{ijl}}$. In its general form the KS field can also be a function of time but we limit
the study to a steady KS.

\subsection{Energy spectrum and Kinematic Simulation}

\begin{figure}[htp]
\begin{center}
\includegraphics [scale=0.8]{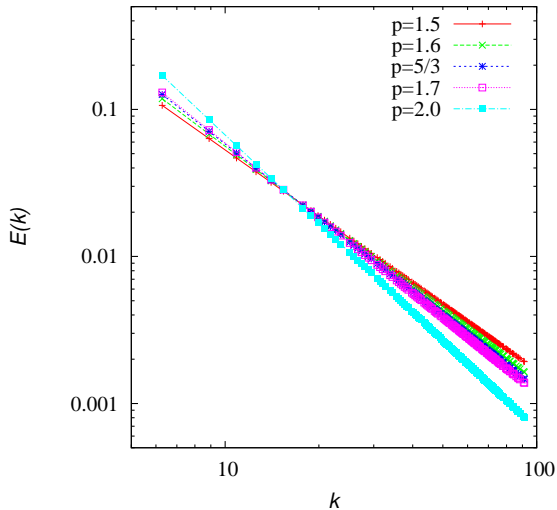}
\end{center}
\caption{\label{F5.0} Spectral energy distribution for different power laws}
\end{figure}

In Kinematic Simulation the underlying Eulerian velocity field is generated as a sum of random incompressible
Fourier modes with a prescribed energy spectrum $E(k)$.
In the present work, the spectrum is chosen as a power
law with an exponent, $p$, varying from 1.4 to 2.5 and is defined as:
\begin{equation}
E(k) = (p-1) \frac{{u}_{0}^2}{k_{min}} \left ( \frac{k}{k_{min}} \right )^{-p}
\label{KSspectra}
\end{equation}
for $k_{min} \leqslant k \leqslant k_{max}$ (see Fig. \ref{F5.0}).
\\
The turbulence rms velocity $u_{rms}$ is fixed so that
the total turbulent kinetic energy density is the same for all cases:
\begin{equation}
\mathcal{E} = \frac{1}{2} \int_{k_{min}}^{k_{max}} E(k) dk \simeq \frac{1}{2} u_{0}^2
\label{KSenergy}
\end{equation}
\\
Tuning the power $p$ allows us to change the energy distribution on scales from
narrow range energetic scales for $p \to 2.5$ to more equi-distribution for $p \to 1.4$.
We expect the Lagrangian attractors' topology to be significantly affected
by the modification in the spectral energy distribution.
The characteristic velocity associated with the smallest scale $\eta$ is given by
\begin{equation}
u(\eta) = \sqrt{ \frac{E(k_{\eta})}{k_{\eta}} }
        = u_0 \sqrt{p-1} \left ( \frac{L}{\eta} \right )^{\frac{1-p}{2}}
\end{equation}
and the associated characteristic time is
\begin{equation}
\tau(\eta) = \frac{\eta}{u(\eta)} = \frac{\eta}{u_0} \frac{1}{\sqrt{p-1}} \left ( \frac{L}{\eta} \right )^{\frac{p-1}{2}}
\end{equation}

\subsection{Periodic KS method for periodic flow}

Following \cite{Farhan-et-al-2015}, the wavevectors
$\mathbf{k}_{ijl}=(k_{i},k_j,k_l)$ follow an arithmetic distribution to enforce a periodic condition
for the velocity flow
field:
\begin{equation}
k_{i} = \frac{2 \pi}{L_{x}} (n_{i}-1), \,\,
k_{j} = \frac{2 \pi}{L_{y}} (n_{j}-1), \,\,
k_{l} = \frac{2 \pi}{L_{z}} (n_{l}-1)
\label{eqn: geometrical decimation1}
\end{equation}
where $(n_{i},n_{j},n_{l})$ are integers satisfying $1 \leqslant n_{i} \leqslant N$. In practice, we choose
$(L_{x}=L_{y}=L_{z})$ for creating an isotropic turbulence and to ensure the flow incompressibility the Fourier
coefficient vectors $\mathbf{a}_{ijl}$ and $\mathbf{b}_{ijl}$ are set orthogonal to the wavevector:
\begin{equation}
\mathbf{a}_{ijl} \cdot \mathbf{k}_{ijl}=\mathbf{b}_{ijl} \cdot \mathbf{k}_{ijl}=0
\end{equation}
Their magnitude is fixed by the energy spectrum, $E(k)$ (\ref{KSspectra}).
\begin{equation}
\left|\mathbf{a}_{ijl}\right|^{2}= \left|\mathbf{b}_{ijl}\right|^{2} = 2E(k)\Delta k_{ijl} / m_k
\label{Fouriercoef}
\end{equation}
where $m_k$ is the number of wavevectors of wavenumber $k=\|\mathbf{k}_{ijl}\|$.
This is the key point for using KS for this study. The use of (\ref{KSspectra}) in (\ref{Fouriercoef}) is straightforward and does not require complicated forcing techniques.
From the spectral law, the rms velocity (\ref{KSenergy}) and the integral length scale can be defined:
\begin{equation}
{\mathcal{L} = { 3 \pi \over 4} { \int_{k_{min}}^{k_{max}} k^{-1} E(k) dk \over \int_{k_{min}}^{k_{max}} E(k)
dk }} \label{a9}
\end{equation}
The Kolmogorov length scale is defined as $ \eta = 2\pi/k_{max}$, whereas the largest physical scale is $L=
2\pi / k_{min}$ which determines the inertial range $[\eta,L]$ over which (\ref{KSspectra})
is observed. It is worth noting that $\mathcal{L} \simeq L$ for sufficiently large inertial ranges. However,
here in contrast to other KS studies the inertial range is small and  $L \simeq 5 \mathcal{L}$. In this paper,
non-dimensional numbers ($St$ and $Fr$) are based on the integral length scale $\mathcal{L}$ and for the sake
of future comparisons both are reported in Table~\ref{tabKS}. The ratio between the largest length scale and
the Kolmogorov length scale is $k_{max} / k_{min}$ and the associated Reynolds number is: $Re_{{L}} =
(k_{max}/k_{min})^{4/3}$. This is the standard way to define a Reynolds number in KS and a DNS or an
experiment yielding the same ratio $k_{max}/k_{min}$ would have a much larger Reynolds number. Finally, a
characteristic time for normalisation can be $t_d = {L}/u_{rms}$ or $\mathcal{T}= \mathcal{L}/u_{rms}$. All
the periodic KS parameters are gathered in Table~\ref{tabKS}.
\begin{table}[h]
\caption{\label{tabKS} Periodic KS parameters}
\newcolumntype{d}[1]{D{.}{\cdot}{#1}}
\begin{tabular}{ld{-1}}
\hline \hline $L_x=L_y=L_z$ & 1
\\
$N$ & 10
\\
$N_p$ & 15625
\\
$u_{rms}$ & 0.8703
\\
$\mathcal{L}$ & 0.2106
\\
$L$ & 1
\\
$\eta$ & 0.0642
\\
$\mathcal{T}$ & 0.2420
\\
$t_d$ & 1.1491
\\
${k_i}_/{k_i}_{min}$ & 9
\\
$k_{max}/k_{min}$ & 15.5885
\\
$Re_L$ & 38.94
\\
\hline \hline
\end{tabular}
\end{table}
\\
The particles are initially homogeneously distributed (this initial distribution is the same for all cases) and whenever a
particle leaves the turbulence box domain (e.g. $\textbf{X}_i >L_x$) it is re-injected from the opposite
side to keep the periodic condition.

\subsection{Equation of motion}

Following \cite{El-Azm-Nicolleau-2008} the equation of motion for the inertial particle is derived from
\cite{Gatignol-1983,Maxey-Riley-1983} and consists of a drag force and drift acceleration (weight):
\begin{equation}
{d{\bf V} \over dt} = \frac{1}{\tau_{a}} \left ( {\bf u}({\bf x}_{p}(t),t)-{\bf V}(t)+{\bf V}_{d} \right )
\label{E2-29}
\end{equation}

\noindent where $\tau_{a}$ is the particle's aerodynamic response time and $V_{d} = \tau_{a}  {\bf g}$ the
particle's terminal fall velocity or drift velocity.

\begin{figure*}[htb]
\begin{center}
\includegraphics [scale=0.32]{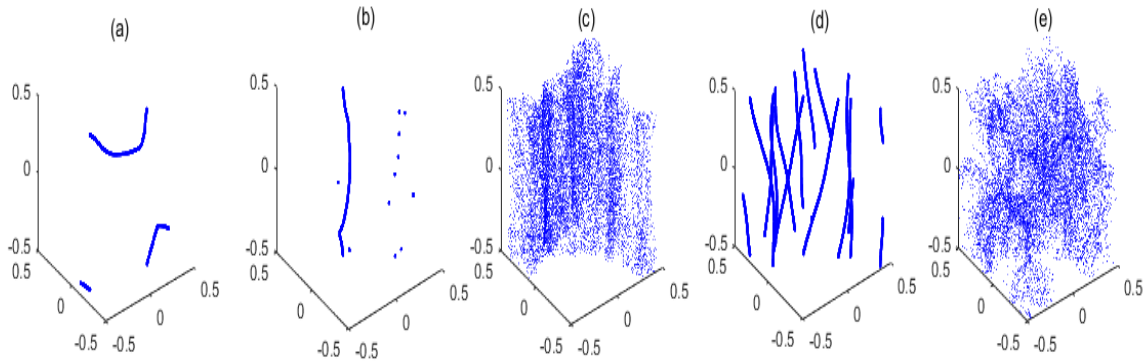}
\end{center}
\caption{\label{Figure2-attracshape}
Different characteristic attractor shapes.}
\end{figure*}

\subsection{Non-dimensional parameters}

Two non-dimensional parameters are introduced to make quantitative analyses of the particle
clustering.

\begin{itemize}
\item
 The Stokes number expresses the ratio between the particle's response time (inertia effect) and the
turbulence characteristic time
    \begin{equation}
        St= \tau_{a}/\mathcal{T} = \tau_{a} u_{rms} / \mathcal{L}
        \label{E2-30}
    \end{equation}
 It measures the relative importance of the particle inertia. In the limiting case $St = 0$; the heavy
particles recover the motion of the fluid tracers, whereas for $St \rightarrow \infty $ the heavy particles
become less and less influenced by the surrounding velocity field.
%
\item
The Froude number is the ratio between inertial forces and gravitational forces.
    \begin{equation}
        Fr = u_{rms}/ \sqrt{g\mathcal{L}}
        \label{E2-31b}
    \end{equation}
In practice, in our study the rms velocity $u_{rms}$ and inertial length scale are constant and $g$ is varied.
\end{itemize}

\section{Clustering patterns, Lagrangian attractors}
\label{sectionIII}

\subsection{Different clustering Patterns}

Figure~\ref{Figure2-attracshape} illustrates the nomenclature we use for the different characteristic shapes
we observed for the Lagrangian attractor.

The particles initially uniformly distributed in the flow field are allowed to evolve until a Lagrangian
attractor is achieved. The shape of the attractor varies from clear one-dimensional structures
to three-dimensional distributed structures.
\begin{itemize}
\item[a)] 1D-H : horizontal one-dimensional Lagrangian attractor as in Fig.~\ref{Figure2-attracshape}a,
\item[b)] 1D-V : vertical one-dimensional Lagrangian attractor as in Fig.~\ref{Figure2-attracshape}b. The
attractor has a point on each top and bottom face of the box
($z=-0.5$ and $z=0.5$).
\item[c)] 2D-L : two-dimensional vertical curtain-like layer as in Fig.~\ref{Figure2-attracshape}c (see also
\cite{Woittiez-et-al-2008}),
\item[d)] 1D-L : Complex 1D layered structure as in Fig.~\ref{Figure2-attracshape}d.
This case was not previously observed for $p=5/3$ and appears only for larger values of $p$.
\item[e)] 3D : Any three-dimensional structure without any particular structure in the cloud as in
Fig.~\ref{Figure2-attracshape}e. This is in fact the most common observation.
\end{itemize}

\subsection{\label{secquant}Quantification of Clustering Patterns - Nearest-neighbour distance analysis}

Visualizations of the particle cloud for small discrete increments of the three parameters $St$,
$Fr$  and $p$ can be tedious. It means looking at thousands of cases in this study in a systematic order. Beyond the
simple visualisation, it is important to quantify the Lagrangian attractors using an appropriate method for
spatial clustering. The average-distance-to-nearest-neighbour is chosen here for direct comparison with
\cite{Farhan-et-al-2015}.
\\[2ex]
The advantage of using this approach is that it is not necessary to reach the final cluster at $t \to \infty$,
a snapshot at earlier times gives us a clear idea of the kind of Lagrangian attractor to expect. The
average-distance-to-the-nearest-neighbour $\Delta$ \cite{Park-Lee-2014} is introduced to systematically
quantify the clustering patterns. At a given time for each particle $\mathbf{X}_m$ its nearest neighbour
is $\mathbf{X}_n=(x_{n},y_{n},z_{n})$.
Then we define the average-distance-to-the-nearest-neighbour as
\begin{equation}
\Delta=\frac{1}{N_p}\sqrt{\sum_{m=1}^{N_p}\Delta^2_{mn}} \label{E2-31d}
\end{equation}
Where
$\Delta^2_{mn}=(x_{m}-x_{n})^{2}+(y_{m}-y_{n})^{2}+( z_{m}-z_{n})^{2} \label{E2-31c}
$.
In practice, the method will detect a one-dimensional structure for $\Delta \leqslant \Delta_{cr1} = 0.008$ while 2D
layered structures are observed for $0.01 = \Delta_{cr2_{-}}\leqslant \Delta \leqslant  \Delta_{cr2_{+}} = 0.014$.
\\[2ex]
We applied the average-distance-to-nearest-neighbour method to all run cases to see the variations in the
attractor patterns for the same time $t=100$ which we found large enough to reach the critical values
$\Delta_{cr1}$, $\Delta_{cr2_{-}}$ and $\Delta_{cr2_{+}}$.

\section{Particle attractors with modified power laws of energy spectrum }
\label{sectionIV}

Before analysing all cases in terms of iso-contours, we first run a few cases for different values of $St$ with and
without ($Fr= \infty$)  gravity  and quantify them using the nearest-neighbour analysis.

\subsection{$0.124 \leqslant St \leqslant 1$}

Initially we consider two values for the Froude number in order to investigate the effect of the spectral power law variations with
increasing values of the Stokes number $St$.
The results with no gravity ($Fr= \infty$) are shown in Fig.~\ref{Figure2}:
\begin{figure}[htb]
\begin{center}
\includegraphics [scale=0.4]{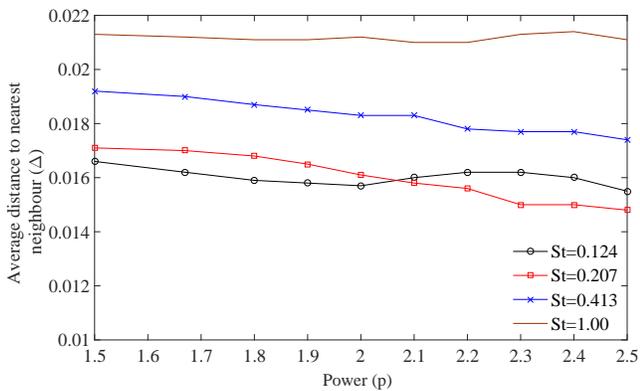}
\end{center}
\caption{\label{Figure2}
$\Delta$ as a function of $p$ for $St=0.124$, 0.207, 0.413 and 1 at $t=100$ and without gravity $Fr=\infty$.}
\end{figure}
$\Delta$ is almost constant so that it can be concluded that
the particles clustering is barely influenced by the spectral power law in the absence of gravity.
\begin{figure}[htb!]
\begin{center}
\includegraphics [scale=0.38]{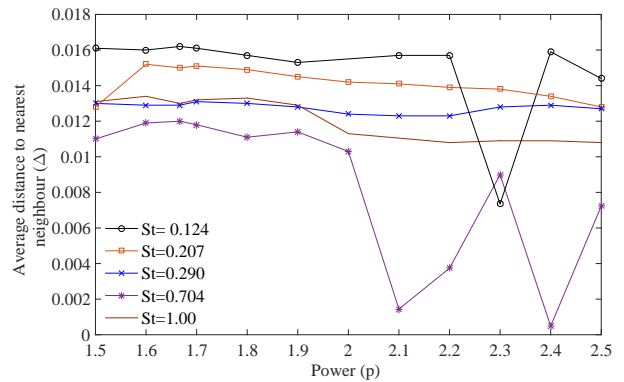}
\caption{\label{Figure4}
$\Delta$ as a function of $p$ for $St=0.124$, 0.207, 0.413 and 1 at $t=100$ and $Fr=0.49$.}
\end{center}
\end{figure}
\\[2ex]
By contrast to the case $Fr=\infty$, the clustering can become more significant when
the Froude number $Fr$ is decreased as shown in Fig.~\ref{Figure4}
for  $Fr=0.49$.
The curves of $\Delta$ as a function of $St$ show troughs characteristics of 1D attractors.
The case $St=1$ seems rather insensitive to that range of $Fr$ numbers and power laws.

\subsection{$0.59 \leqslant Fr \leqslant 1.2$}

\begin{figure}[htb]
\begin{center}
\includegraphics [scale=0.4]{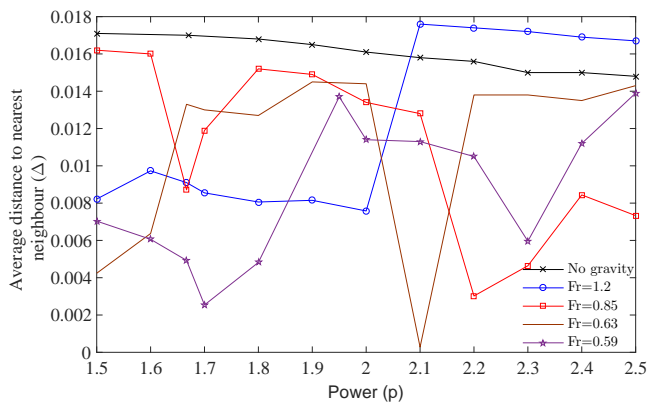}
\end{center}
\caption{\label{Figure5}
$\Delta$ as a function of $p$ for $Fr=0.57$, 0.63, 0.85, 1.2 and $\infty$ at $t=100$;  $St= 0.207$.}
\end{figure}

\noindent It is equally important to observe the clustering variations for some cases with a constant Stokes
number by varying the Froude numbers and the power laws.
Fig.~\ref{Figure5} shows the case $St=0.207$ for $Fr \in [0.57,\infty[$.
\\
It confirms the previous result that in the absence of gravity ($Fr=\infty$) the energy distribution has little effect on
the clustering pattern. Apart from that result there is no particular trend when varying the Froude number but there is a clear
effect of the power law on all cases with gravity ($Fr \neq \infty$) where different troughs characteristics of 1D Lagrangian attractors can be observed.

\section{Comprehensive analysis with power law variations}
\label{sectionV}

Previous results give a useful but limited insight of the effect of the spectral law,
we now analyse the clustering of inertial particles with different power laws of energy spectrum
fixing either $St$ or $Fr$ and varying the two other parameters.

Iso-contours of $\Delta$ are plotted as functions of
($St$, $p$) for a given $Fr$ (Figs \ref{Figure6-Fr0.89}, \ref{F5.2} and \ref{F5.3}) in Sec.~\ref{contFr} or as functions of ($Fr$, $p$) for a given $St$ in Sec.~\ref{contSt} (Figs \ref{F5.5}, \ref{F5.6} and \ref{F5.7}).
Results are summarised in Tables~\ref{Table-5.1} and ~\ref{Table-5.2} respectively.

Colour-wise light gray (blue online) corresponds to 1D Lagrangian attractor,
very light gray - around 0.012-0.014 (yellow-green online) to the 2D-L and dark gray (dark red online) to 3D structures.
In all the graphs, typical attractors are represented in rows (b-e):
row (b) corresponds to $p \in [2.4, 2.5]$,
row(c) to $p \in [2, 2.1]$,
row(d) to the reference case (Kolmogorov spectrum) $p=5/3$,
and row (e) to  $p=1.5$.
\\[2ex]
The cases $p=5/3$ are hereinafter referred to as `standard' or `reference' case as we can analyse the departure from the attractor found for $p=5/3$ when we vary $p$.

\subsection{\label{contFr} Analysis in relation to constant $Fr$}

Cases for different Froude numbers $Fr$  are listed in Table
~\ref{Table-5.1}. The
power law exponent is in the range $1.4 \leqslant p \leqslant 2.5$. Each case  has been quantified using the nearest neighbour analysis for varying values of the
Stokes number, $0.041 \leqslant St \leqslant 1$. We run almost 300 cases for each $Fr$  and the iso-contours of the
nearest-neighbour distance $\Delta$ as a function of  $(St,p)$ are plotted for three representative cases in Fig.~\ref{Figure6-Fr0.89}a-
Fig.~\ref{F5.3}a. Characteristic 3D-plots of the cloud are also shown to visualize the variations in the
attractors with respect to the power law exponents $p$.

\subsubsection{{`High' $Fr$ values: 0.89 and 0.59}}

\begin{figure*}[htp!]
\begin{center}
\includegraphics [scale=0.23]{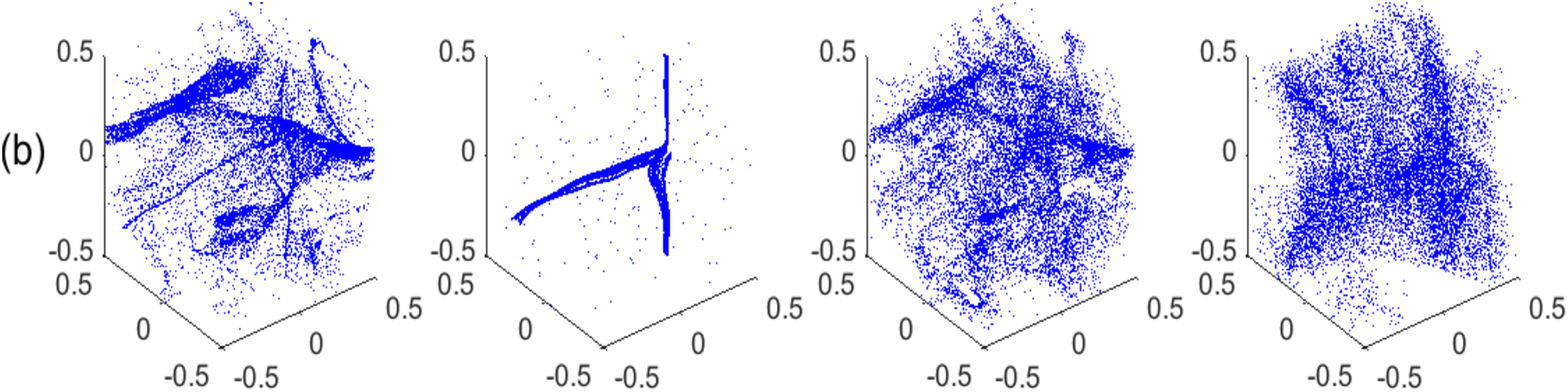}
\includegraphics [scale=0.23]{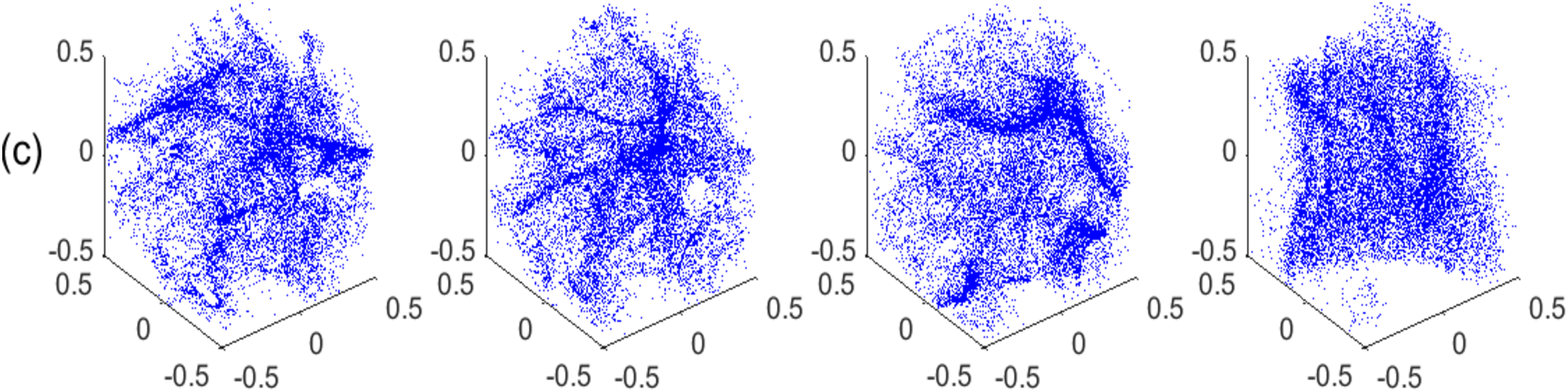}
\includegraphics [scale=0.75]{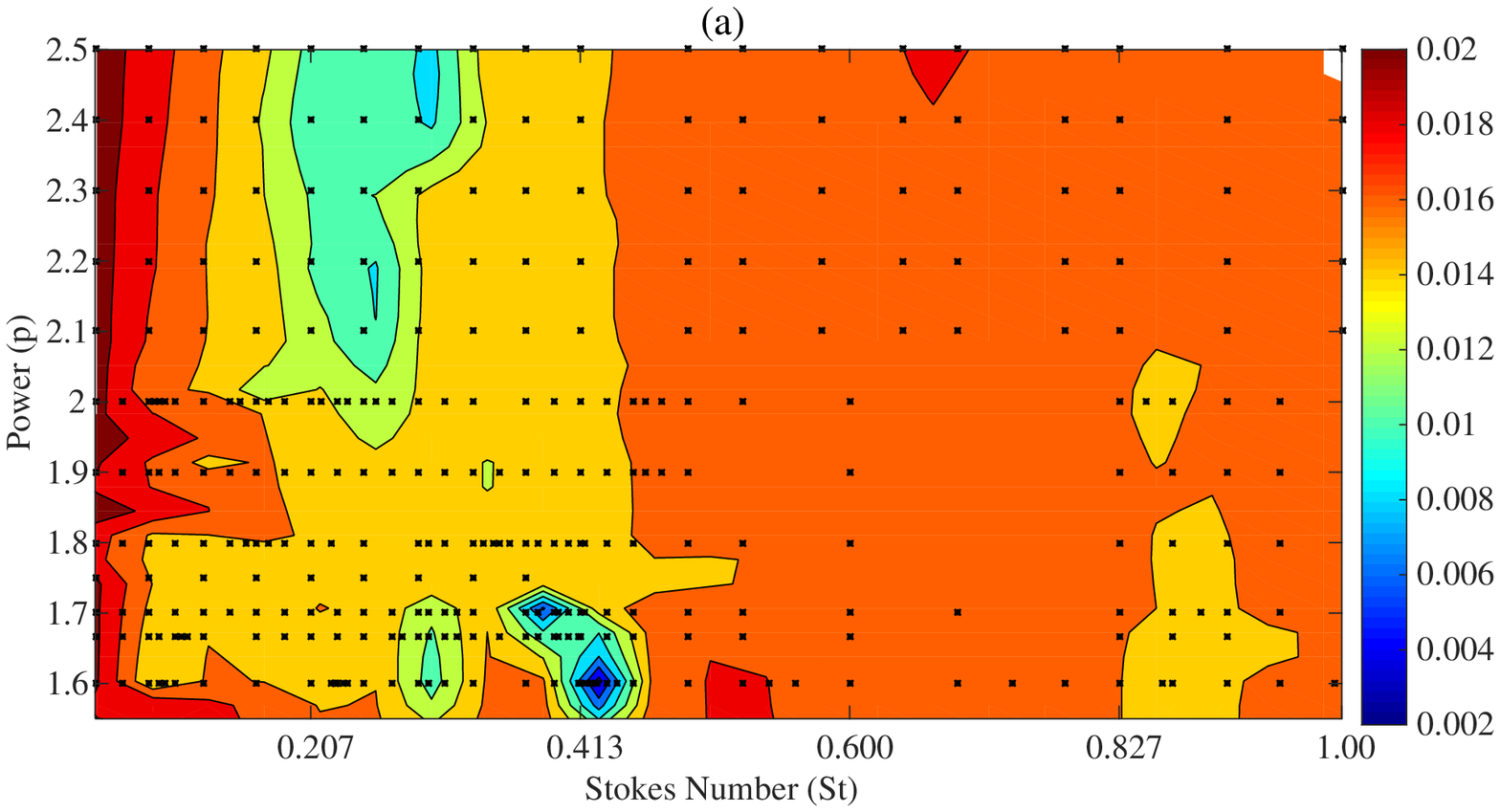}
\includegraphics [scale=0.23]{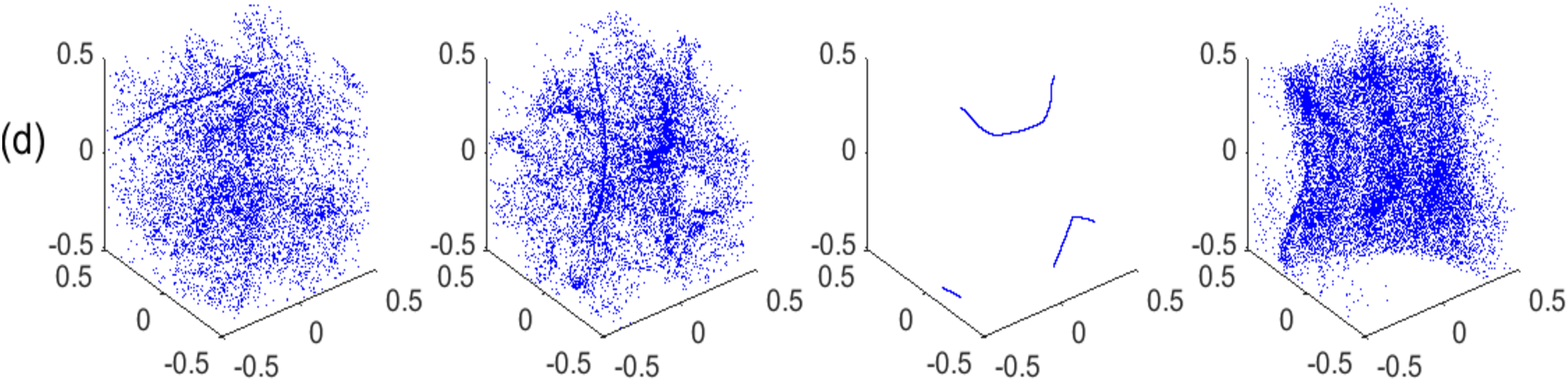}
\includegraphics [scale=0.23]{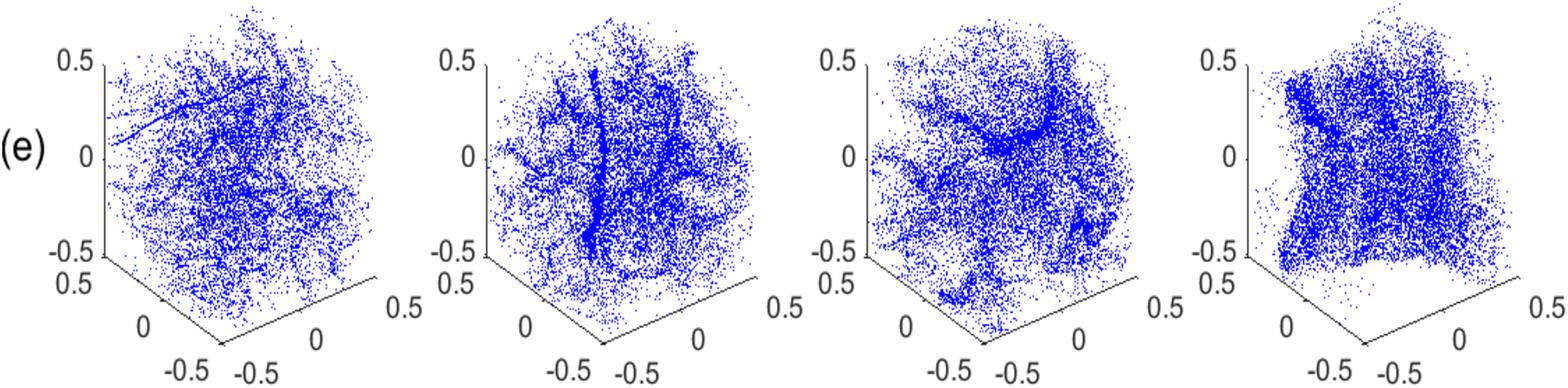}
\end{center}
\caption{\label{Figure6-Fr0.89} 
$Fr=0.89$:
(a)- Iso-contours of $\Delta$ as functions of
$(St,p)$.
Particle clusters with different power laws of energy spectrum
for increasing values of $St$.
(b) $p=2.5$,
(c) $p=2.1$,
(d) $p=5/3$ and
(e) $p=1.5$
and from let to right $St=0.09$, 0.298, 0.413 and 0.91.
}
\end{figure*}

Figure~\ref{Figure6-Fr0.89} shows the cases for $Fr=0.89$.
A well-defined 1D attractor is observed around $St=0.413$
(row (d)) for $p=5/3$.
As $p$ departs from $5/3$ this typical 1D attractor is lost.

None are observed
for $p=1.5$ but 1D attractors reappear for steeper spectral laws ($p \geqslant 2.1$) at lower Stokes numbers
($St \in [0.207,0.400]$) but over a larger range of Stokes numbers.

The attractors also reappear in different shapes. For example, we observe a 1D-H attractor for $St=0.413$ with $p=5/3$, but this 1D-H
attractor is reshaped into a different 1D-V attractor with $p=2.5$ as illustrated in Fig.~\ref{Figure6-Fr0.89}
row(b) ($St=0.298$). Therefore, increases in the power law not only affect the value of the Stokes number $St$ at which a
one-dimensional attractor appears but can also change the orientation and shape of the attractor.
Alteration in the attractor shape can be expected as the turbulence energy is redistributed over different Eulerian structures
but the change from 1D-H to 1D-V is significative as it is an indication that the gravity effect may be enhanced by the
Eulerian velocity field topology - that in KS is governed by the spectral law.

By contrast, the 2D layered attractor observed for large Stokes numbers ($St=0.91$) is fairly independent of the
spectral law so that it is merely a function of $St$ and $Fr$ and the Eulerian velocity field topology has no effect on it.
\\[2ex]
\begin{figure*}[htp]
\begin{center}
\includegraphics [scale=0.23]{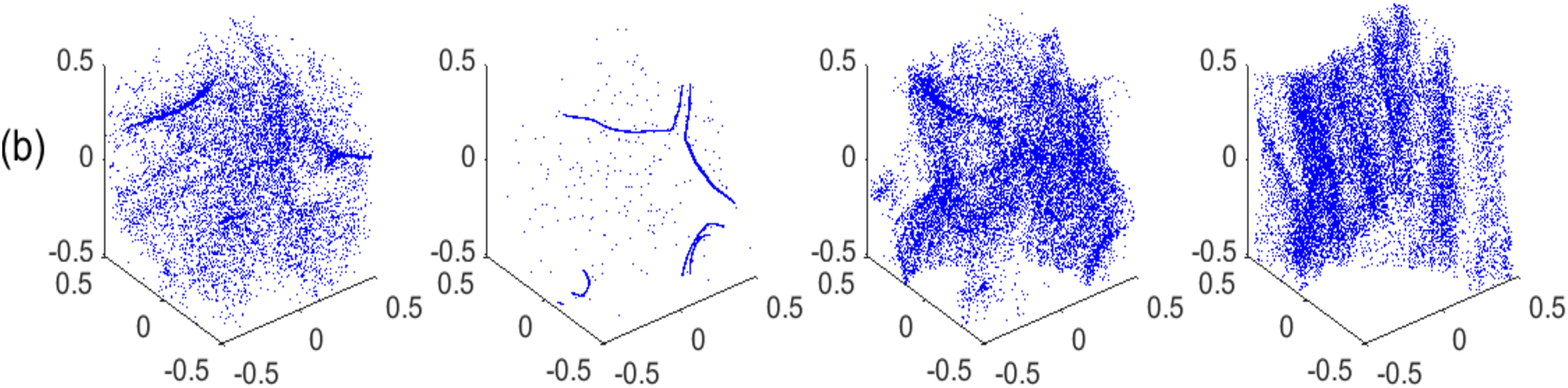}
\includegraphics [scale=0.23]{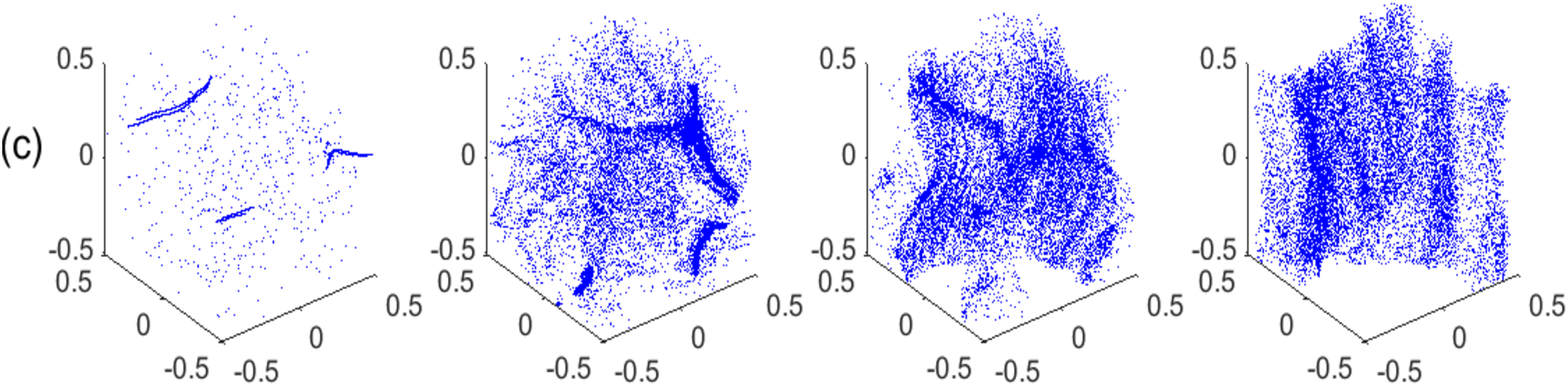}
\includegraphics [scale=0.75]{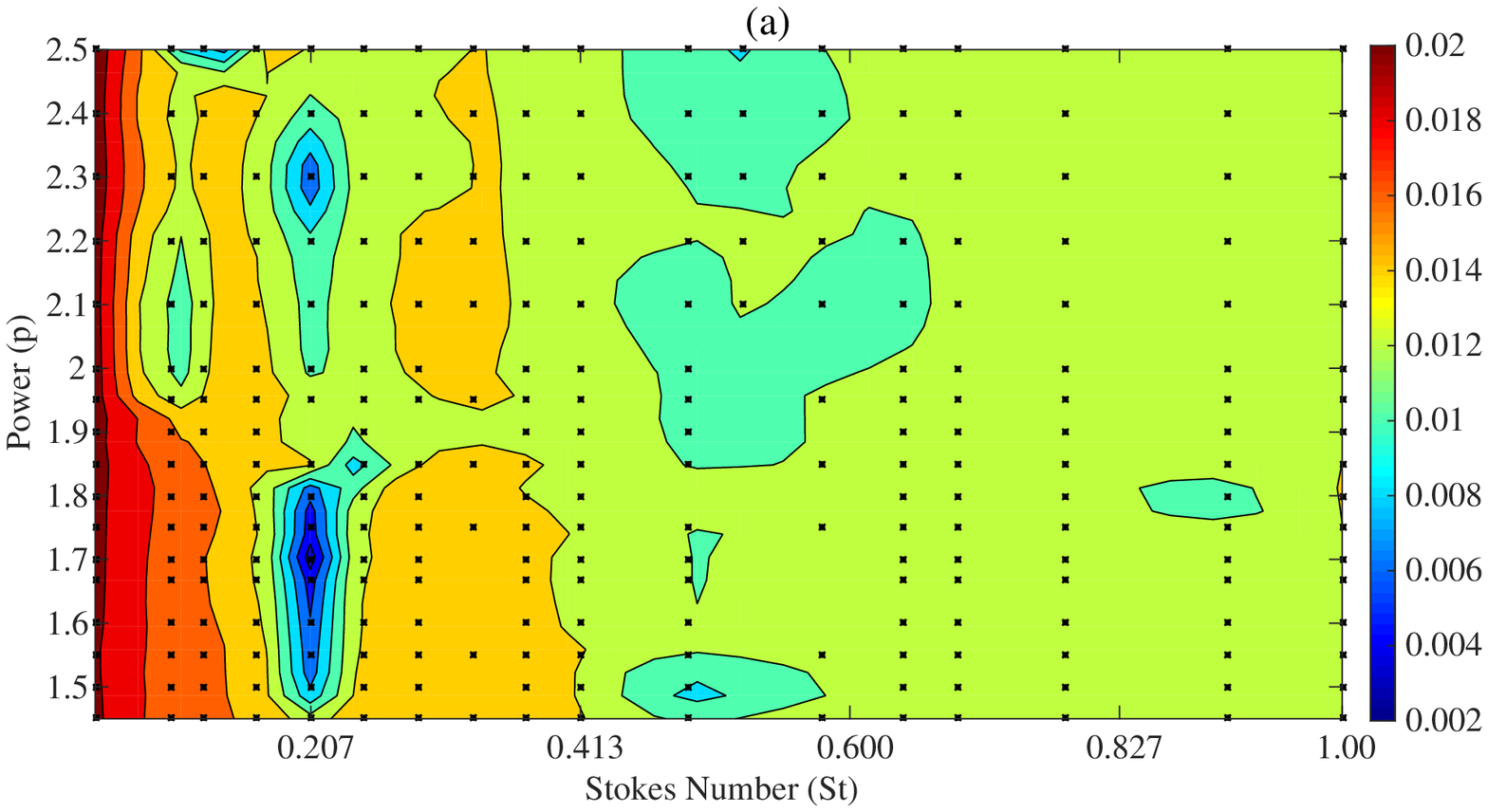}
\includegraphics [scale=0.23]{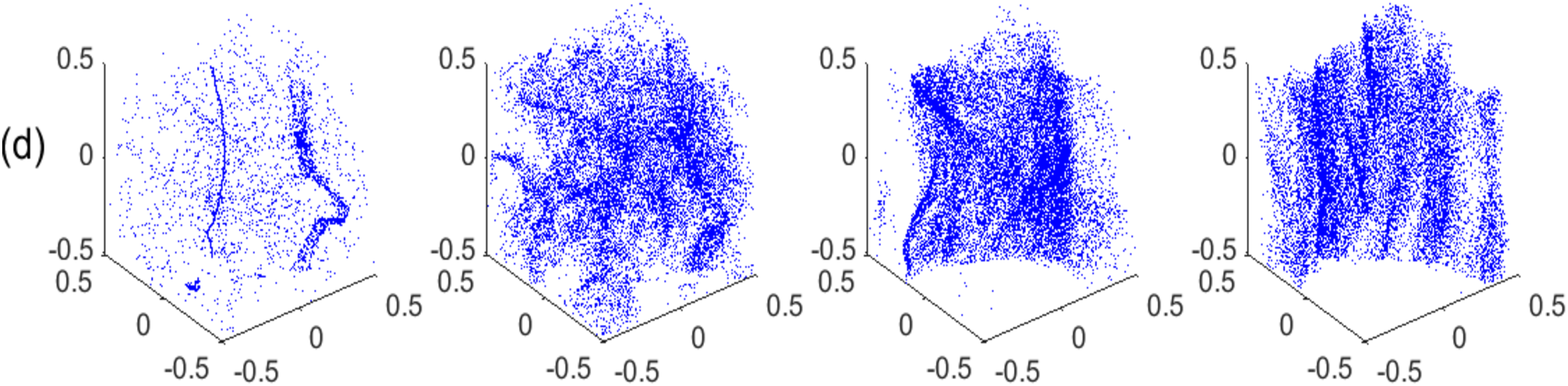}
\includegraphics [scale=0.23]{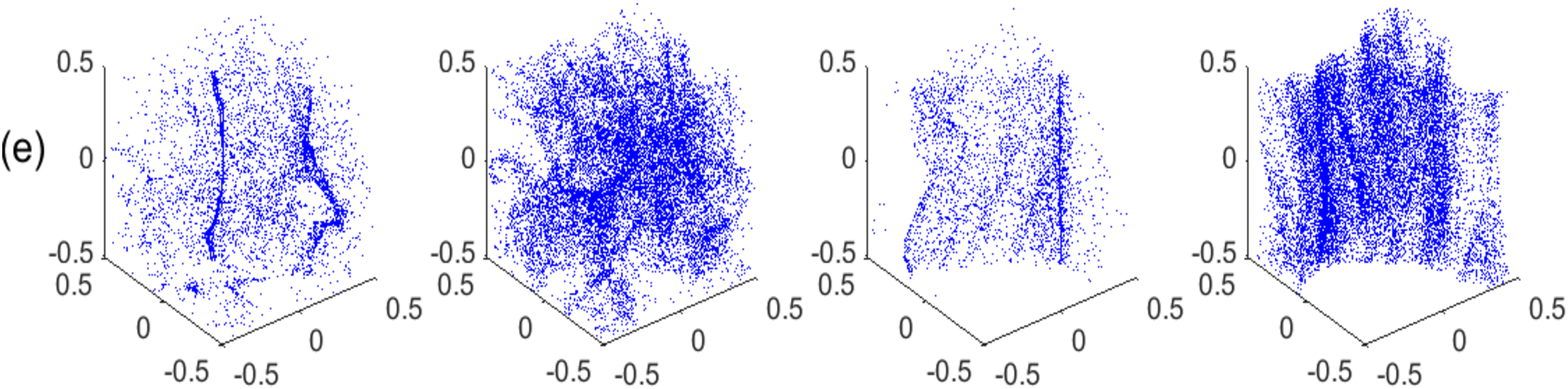}
\end{center}
\caption{\label{F5.2}
$Fr=0.59$
(a)- Iso-contours of $\Delta$ as functions of
$(St,p)$.
Particle clusters with different power laws of energy spectrum
for increasing values of $St$.
(b) $p=2.4$,
(c) $p=2.1$,
(d) $p=5/3$ and
(e) $p=1.5$
and from let to right $St=0.1$, 0.207, 0.413 and 0.91
}
\end{figure*}
The second value of the Froude number is $Fr=0.59$ and the
results are plotted in Fig.~\ref{F5.2}. For this $Fr$ value,
the reference case with $p=5/3$ is a 1D-V attractor where the particles accumulate for $St=0.207$
(Fig.~\ref{F5.2} row (d)).
This attractor remains for a larger range of power law ($p \in [1.5,1.8]$) it even reappears for
steeper energy distributions $p \in [2.2,2.4]$ but as a 1D-H attractor.

So for this lower value for $Fr$ it is still possible to affect the attractor's topology by
varying the spectral power law.
A new 1D-H attractor appears for $St=0.1$ and $p \in [2,2.2]$
as shown in Fig.~\ref{F5.2} row (c).
These variations in 1D attractor orientation
show the effect of the power law variations in relation to $Fr$.

The result observed for $Fr=0.89$ for the 2D layered attractors is confirmed, that is
the 2D layered attractor observed for large Stokes numbers ($St=0.91$) is fairly independent of the
spectral law.

By
comparing both cases $Fr=0.89$ and 0.59, we can deduce that the Lagrangian
attractor topology is more immune to variation of $p$ for lower values of $Fr$.

We can also conclude comparing results for both Froude numbers that, in accordance with previous results for $p=5/3$ \cite{Farhan-et-al-2015}, the decrease in the Froude number allows for the formation of the layers at lower Stokes numbers (here $St \geqslant 0.413$). This result is independent of the spectral power law we chose.

\subsubsection{{`Low' $Fr$ values: 0.49 and 0.32}}

\begin{figure*}[htp]
\begin{center}
\includegraphics [scale=0.23]{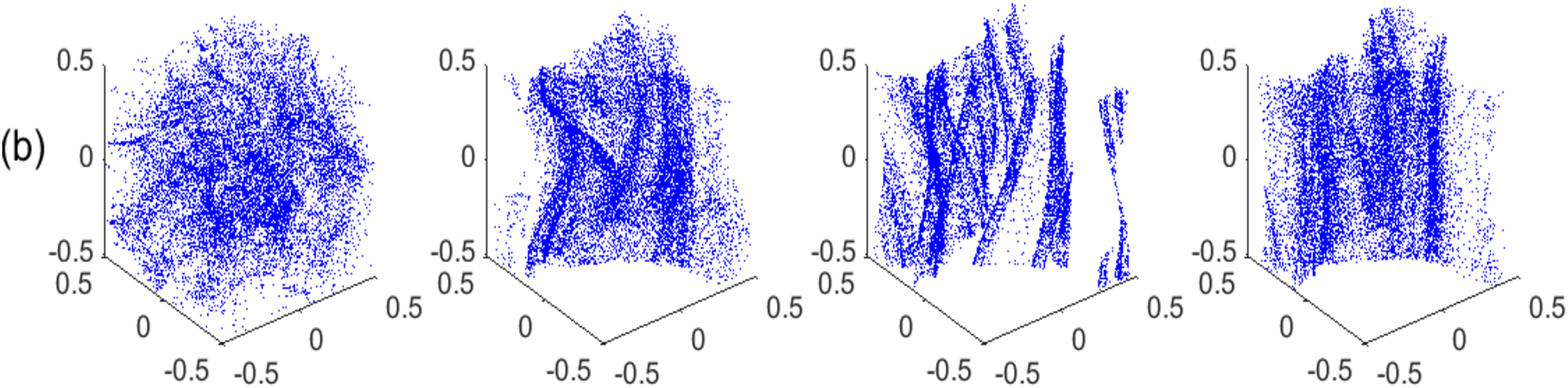}
\includegraphics [scale=0.23]{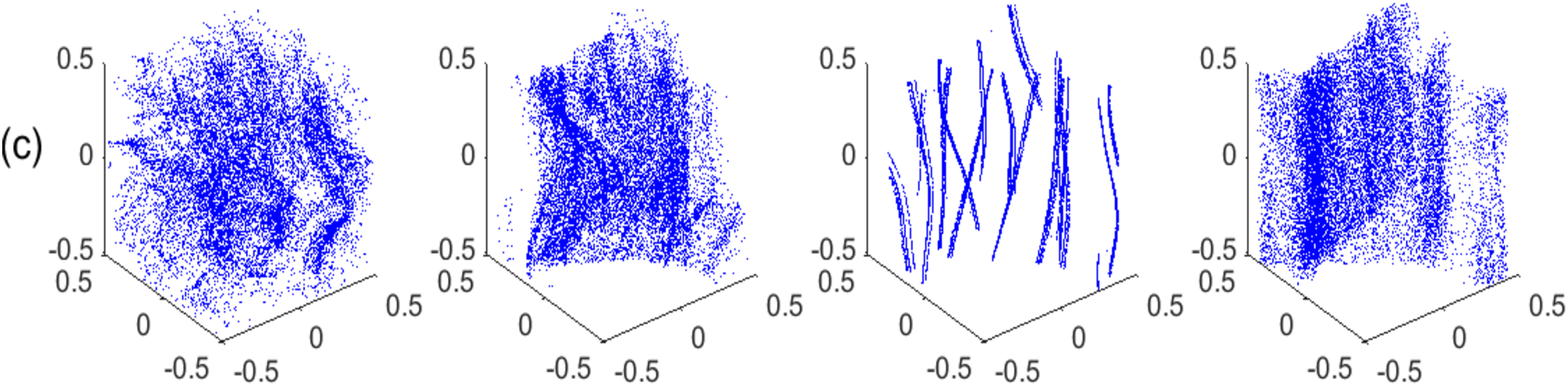}
\includegraphics [scale=0.75]{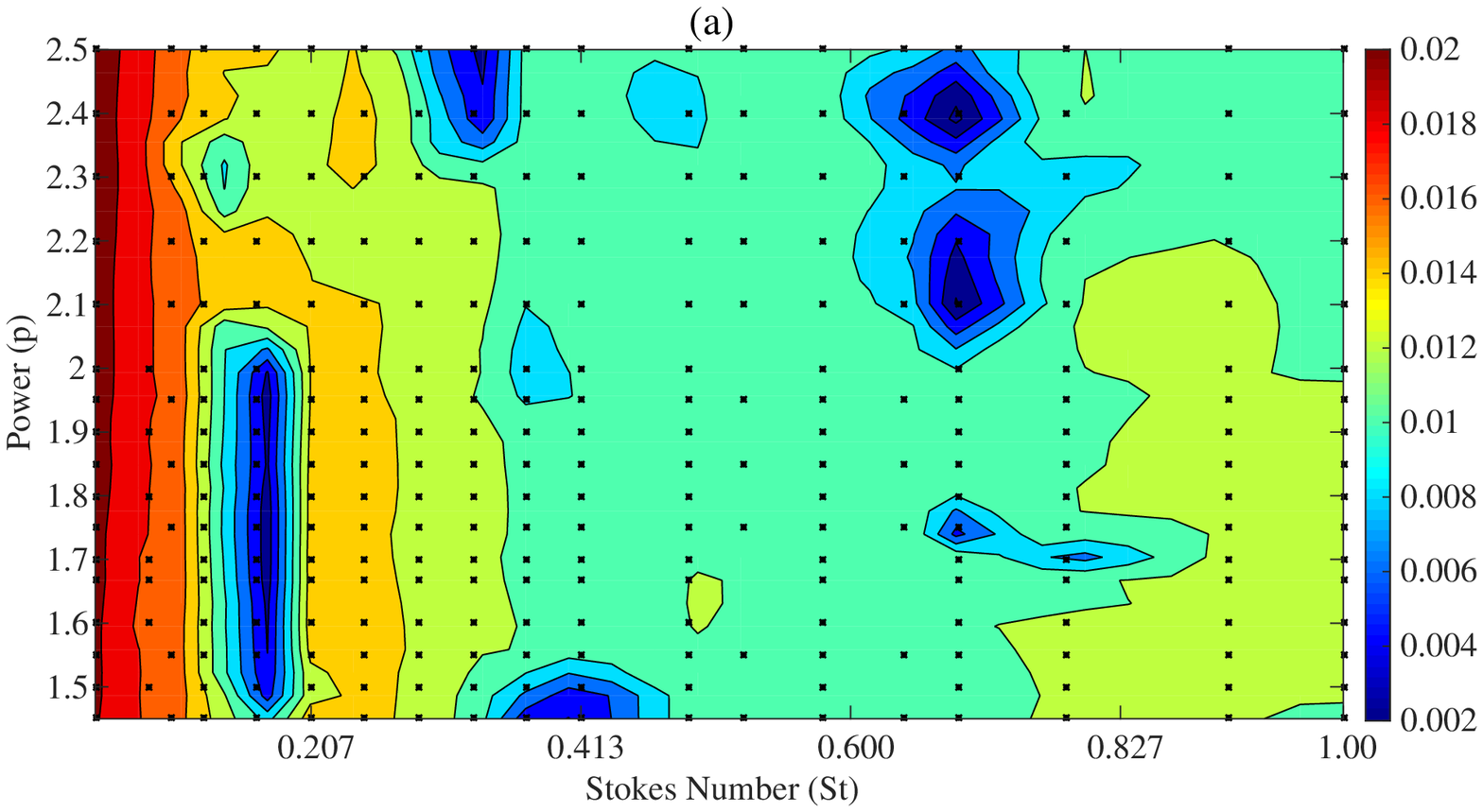}
\includegraphics [scale=0.23]{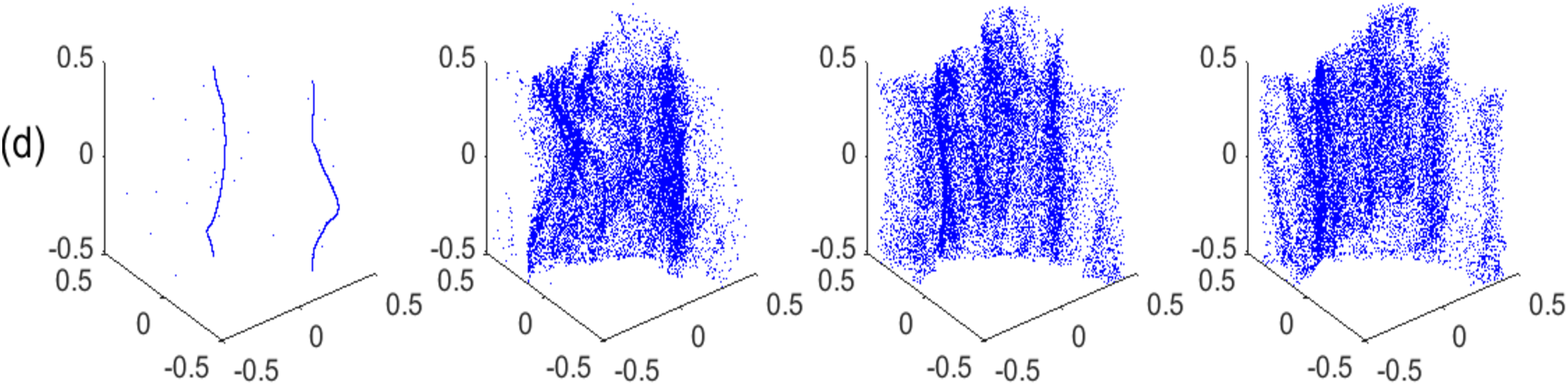}
\includegraphics [scale=0.23]{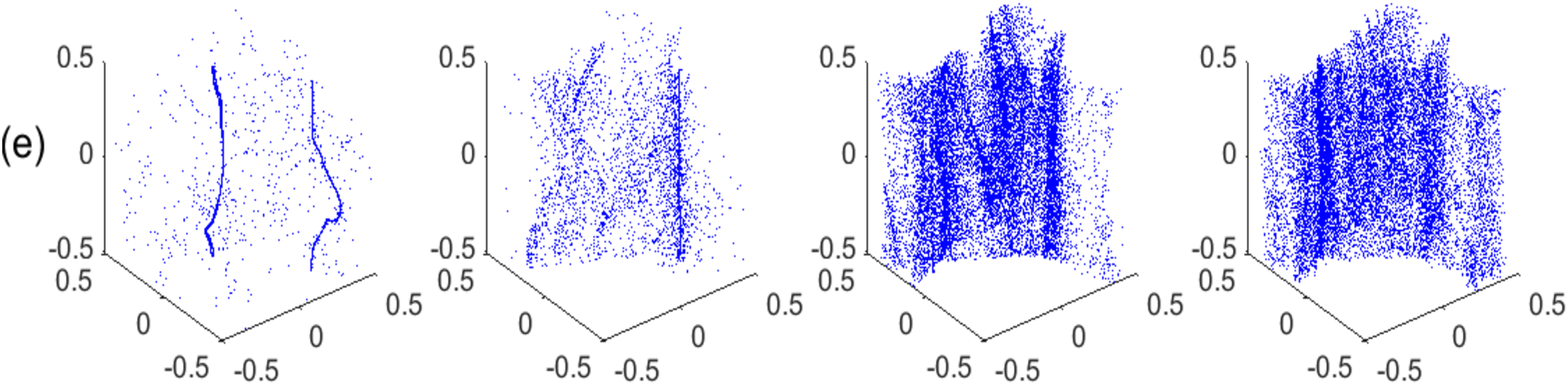}
\end{center}
\caption{\label{F5.3}
$Fr=0.49$:
(a)- $\Delta$ as a function of $(St,p)$.
Particle clusters with different power laws of energy spectrum
for increasing values of $St$.
(b) $p=2.5$,
(c) $p=2.1$,
(d) $p=5/3$ and
(e) $p=1.5$
and from let to right $St=0.16$, 0.413, 0.70 and 0.91}
\end{figure*}

\noindent We carry on the analysis decreasing further the Froude number to $Fr=0.49$ (Fig.~\ref{F5.3}). The Kolmogorov energy spectrum for this particular value of  $Fr$ generates a 1D-V attractor
with $St=0.16$ as shown in  Fig.~\ref{F5.3} row(d).
This attractor is particularly resilient to changes in the spectral law as it persists for
$p \in ]1.5,2]$.
That is in line with previous observations indicating that with higher gravity effects the
1D-V attractor is observed over a larger range of spectral energy distributions.

The 2D layered attractors are still to be observed for $St \geqslant 0.413$ but are somehow
re-inforced showing an inhomegeneous distribution of particles and clustering concentration within the layer itself.
In extreme cases as ($St=0.7, p=2.1$) the 2D layer is shredded into multi-1D-H attractors giving rise
to a new attractor type that we labelled complex 1D layered (1D-L).

It is also found that the range of Stokes numbers for which a 1-D attractor  can appear
increases with decreasing values of $Fr$. For instance, this range is [0.2-0.42] for $Fr=0.89$; while for
$Fr=0.49$, it expands to [0.16-0.70].
\\[2ex]
We can conclude this section by summarising all of our results in Table~\ref{Table-5.1} which also includes cases for $Fr=0.32$.
\begin{table}[h]
\caption{\label{Table-5.1} Occurrences of attractors for given $Fr$ numbers varying $St$ and $p$.}
\begin{tabular}{|c|c|c|c|c|c|}
\hline \hline
\multirow{2}{*}{{Attractor}}  &  &
\multicolumn{4}{c|}{{Froude number $Fr$}}
\\ \cline{3-6}
                                     &                                          & {0.89}       &
{0.59}       & {0.49}       & {0.32}       \\ \hline \hline
\multirow{2}{*}{{1D-H}}       & $St$                     & $<0.5$        &
$<0.25$        & \multirow{2}{*}{No} & \multirow{2}{*}{No}
\\ \cline{2-4}
                                     & $p$                       & 1.5-2.5              &
2.1-2.5             &                     &                     \\ \hline \hline
\multirow{2}{*}{{1D-V}}       & $St$                      & 0.2-0.3             &
0.2-0.3             & 0.1-0.4             & 0.1-0.3             \\ \cline{2-6}
                                     & $p$                      & 1.5-2            &
1.5-1.8             & 1.5-2.5             & 1.5-2.5             \\ \hline \hline
\multirow{2}{*}{2D-L}       & $St$                      & $>0.5$     &
$>0.3$     & $>0.2$     & $>0.1$     \\ \cline{2-6}
                                     & $p$                       & all             &
all             & all             & all             \\
\hline \hline
\multirow{2}{*}{1D-L} & $St$                      & \multirow{2}{*}{No} &
\multirow{2}{*}{No} & 0.4-0.85            & 0.1-0.3             \\ \cline{2-2} \cline{5-6}
                                     & $p$                       &                     &
& 1.7-2.5             & 1.7-2            \\ \hline \hline
\end{tabular}
\end{table}

\subsection{\label{contSt} Analysis in terms of constant $St$}

In this section, we fix the Stokes number $St$ and the changes in attractors' patterns are analysed in terms of varying the values of
$p$ and $Fr$.  We examine the variations in clustering for four different values of the Stokes number namely 0.124, 0.207, 0.413 and 1.
The Froude number varies in the range $0.49 \leqslant Fr \leqslant 1.34$
and $p \in [1.5,2.5]$.

\subsubsection{`Low' $St$ values: 0.207}

\begin{figure*}[htp]
\begin{center}
\includegraphics [scale=0.23]{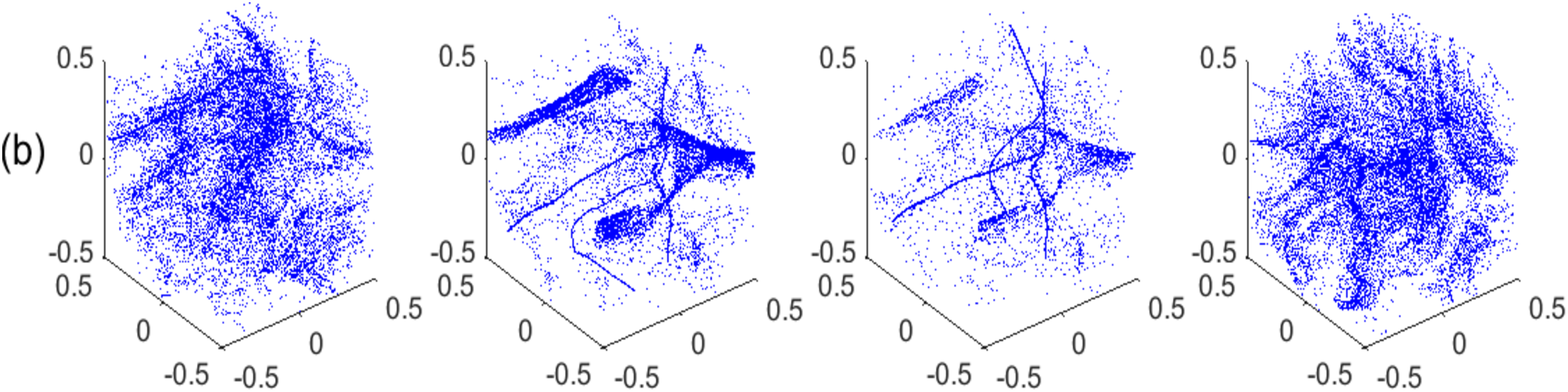}
\includegraphics [scale=0.23]{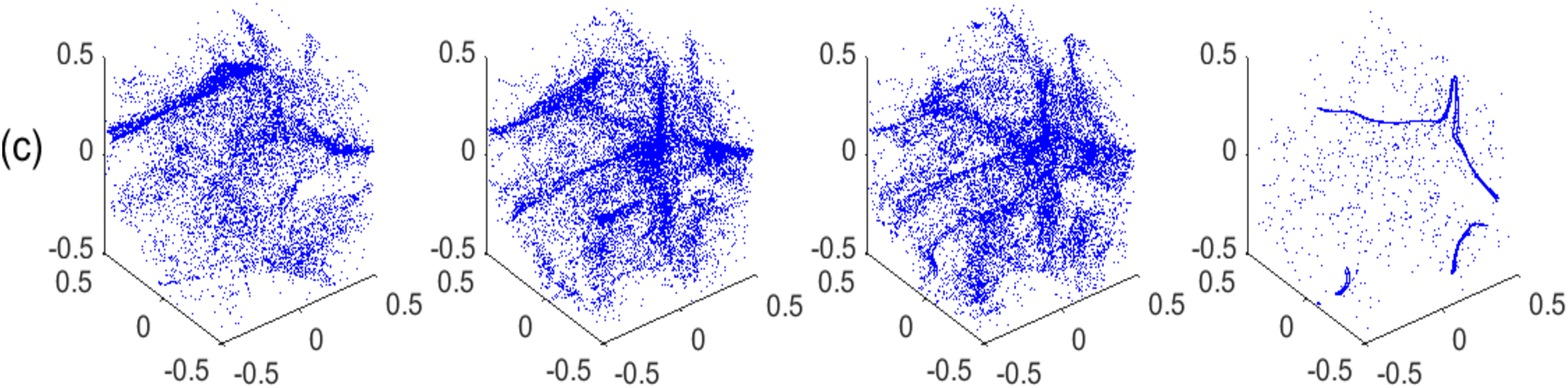}
\includegraphics [scale=0.75]{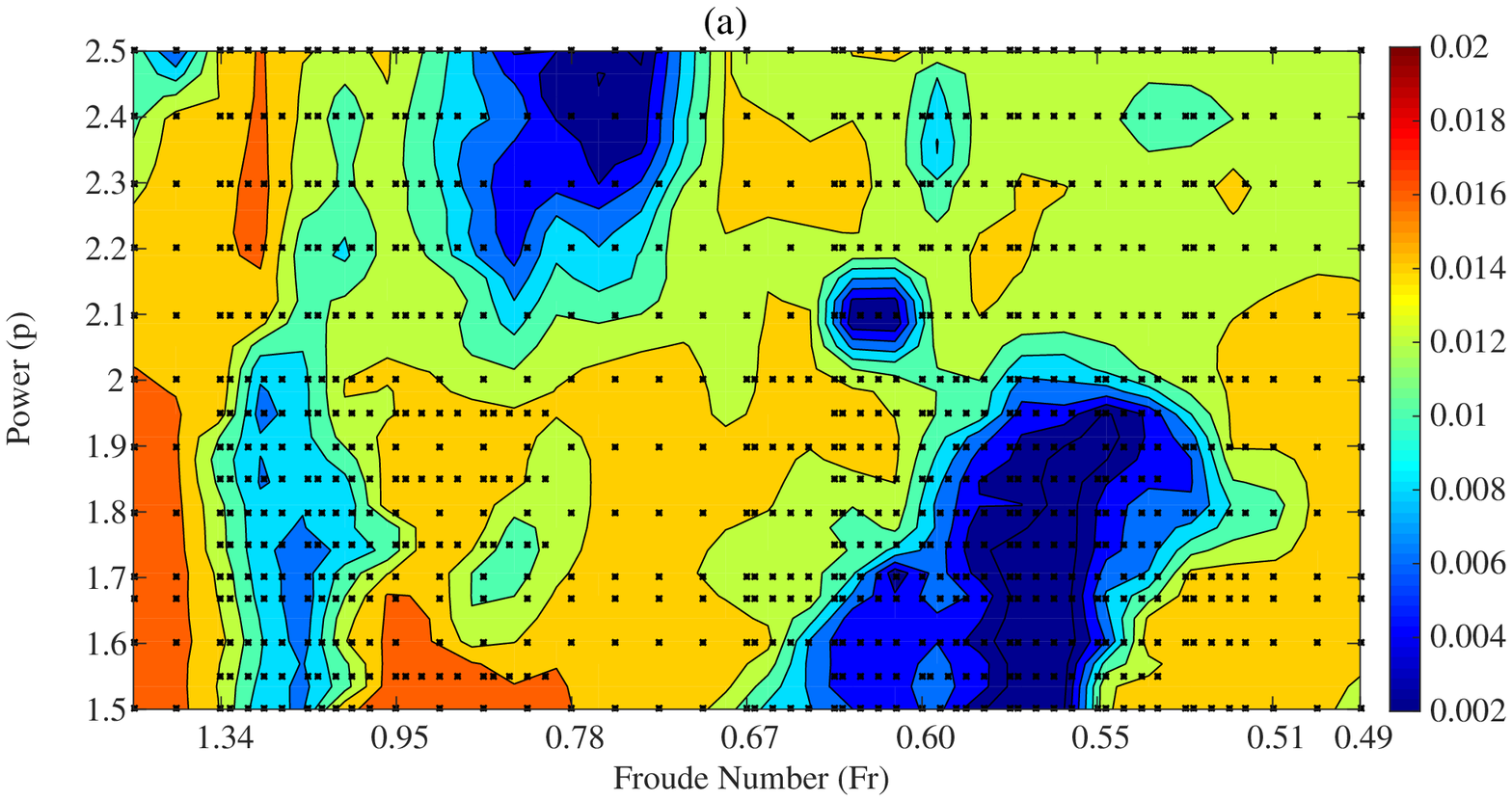}
\includegraphics [scale=0.23]{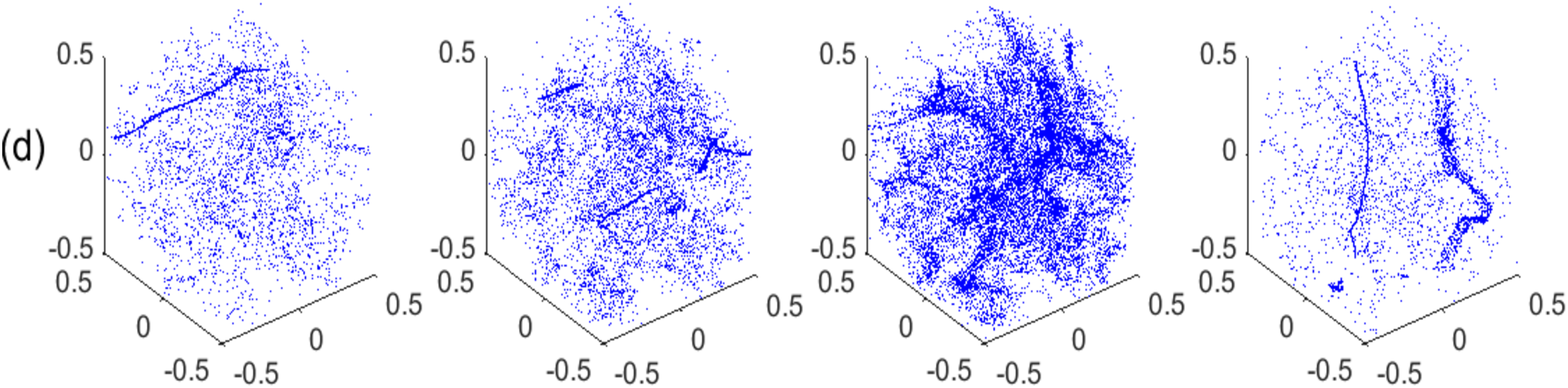}
\includegraphics [scale=0.23]{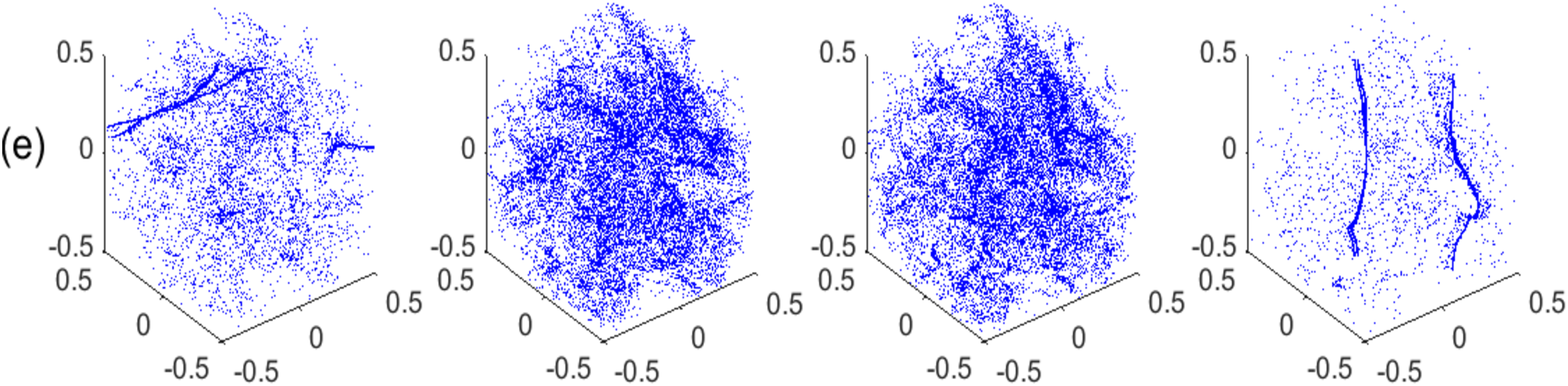}
\caption{\label{F5.5}
$St=0.207$:
(a)- Iso-contours of $\Delta$ as functions of
$(Fr,p)$. Typical particle clusters for different spectral power laws and
decreasing values of $Fr$:
(b) $p=2.5$,
(c) $p=2.1$,
(d) $p=5/3$ and
(e) $p=1.5$
and from let to right $Fr=1.1$, 0.85, 0.72 and 0.6}
\end{center}
\end{figure*}

\noindent For a low value of the Stokes number ($St=0.207$, Fig.~\ref{F5.5}), three different types of 1-D attractor appear.
A 1D-H attractor appears for the high value of  $Fr=1.1$ and this  horizontal attractor persists for $p \in [1.5,2]$ (1st column, row c, d and e). For higher values of $p$ ($p=2.5$) the 1D-H is recovered
but at a higher Froude number around $Fr=1.4$.
\\[2ex]
\noindent For a mid-ranged value of $Fr$ ($[0.7,1]$), a 1D-H attractor appears which reshapes into  more
complex structures with $p>2$ as seen in Fig.~\ref{F5.5} rows (b) and (c).
For low values of $Fr$,
in
Fig.~\ref{F5.5}(a), the light gray (blue online) area  between $0.65<Fr<0.5$ represents a strong one-dimensional clustering.
This attractor shape is retained by
the cluster even when $p$ is varied as evidenced from the contour plot in Fig.~\ref{F5.5}a.
This area stretches backward and forward for respectively decreasing and increasing values of the power law $p$. This means that for $St=0.207$ in the range $0.65<Fr<0.5$ in order to achieve a one-dimensional attractor, one has to
increase the gravity effect when the power law is increased and vice versa.

\subsubsection{`High' $St$ values: 0.413 and 1}
\begin{figure*}
\begin{center}
\includegraphics [scale=0.23]{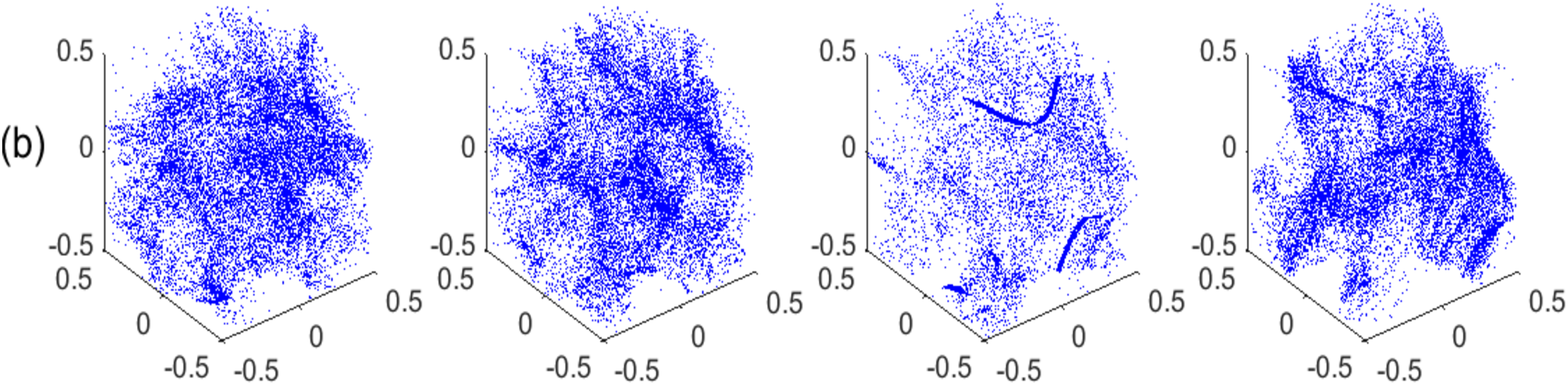}
\includegraphics [scale=0.23]{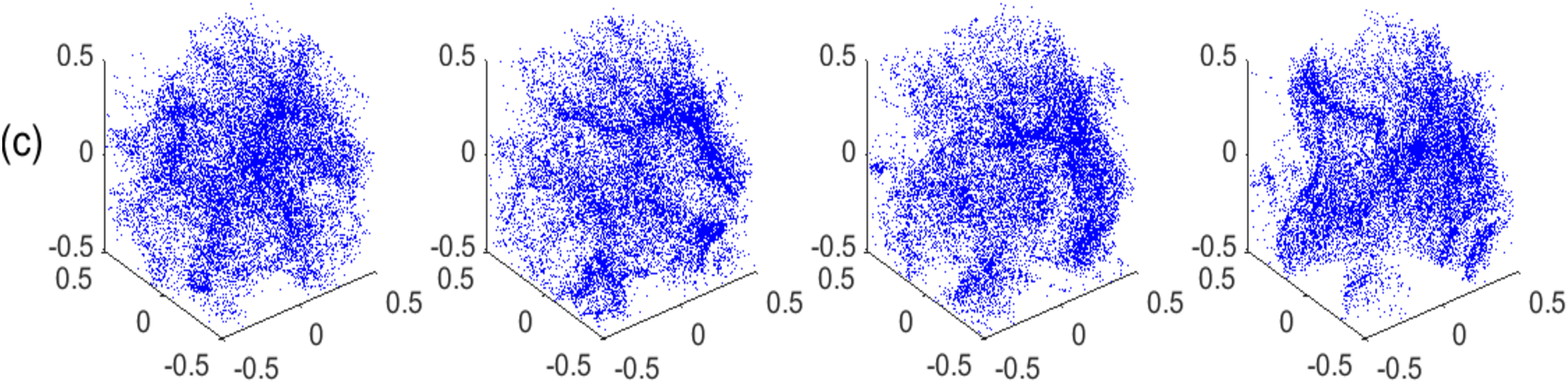}
\includegraphics [scale=0.75]{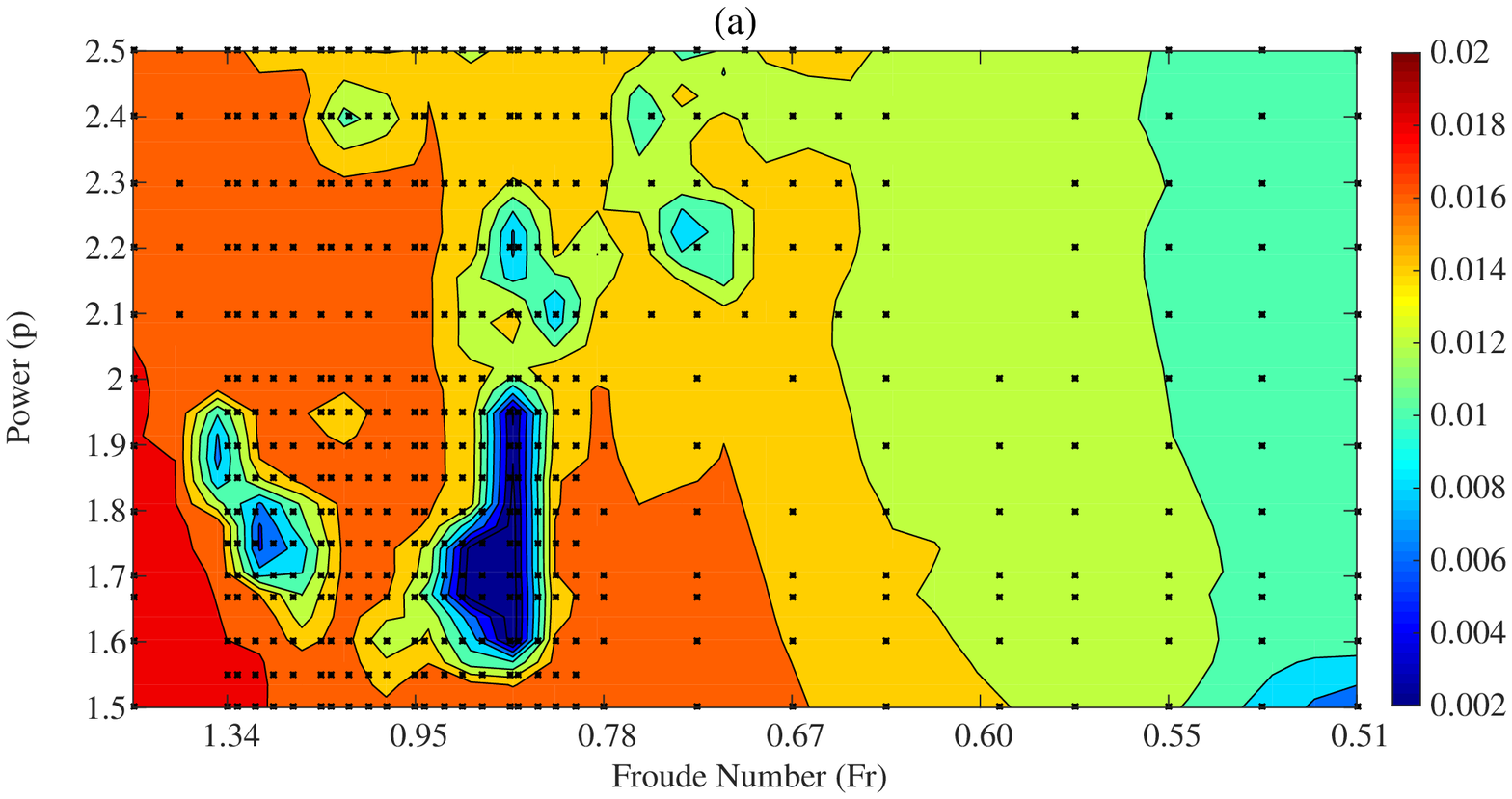}
\includegraphics [scale=0.23]{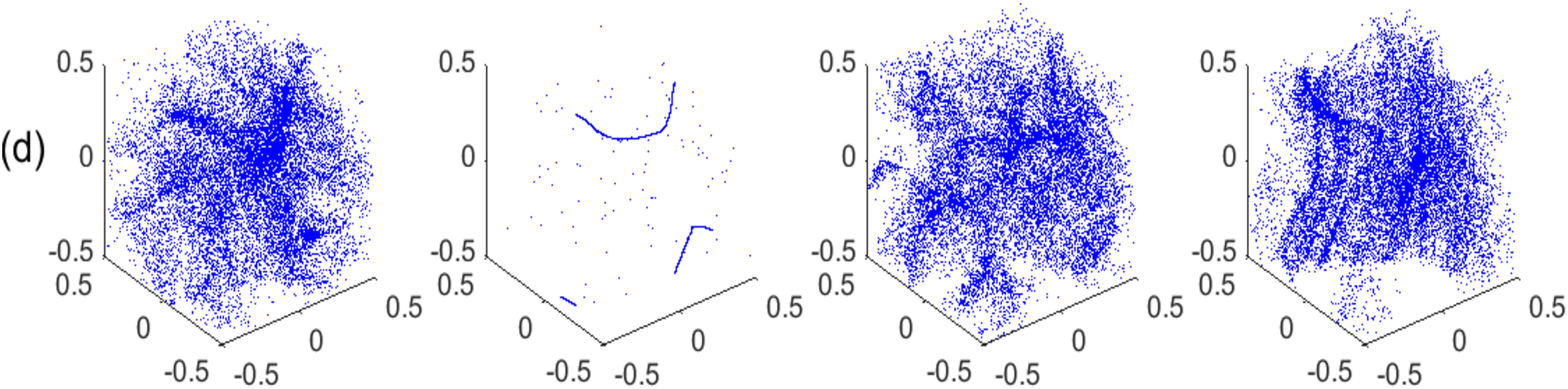}
\includegraphics [scale=0.23]{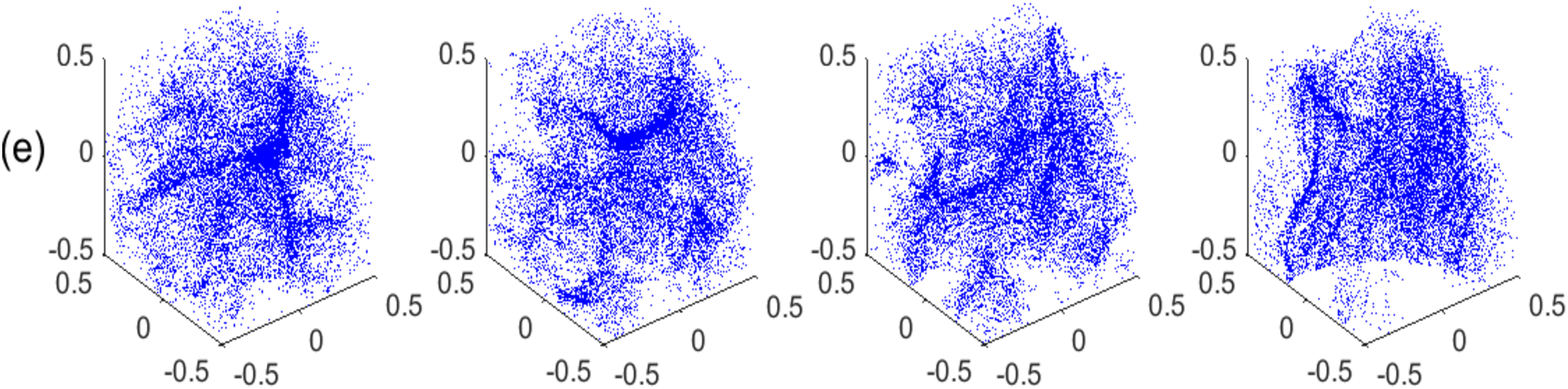}
\caption{\label{F5.6}(
$St=0.413$:
(a)- Iso-contours of $\Delta$ as functions of
$(Fr,p)$. Typical particle clusters for different spectral power laws and
decreasing values of $Fr$:
(b) $p=2.5$,
(c) $p=2.1$,
(d) $p=5/3$ and
(e) $p=1.5$
and from let to right $Fr=1.1$, 0.85, 0.72 and 0.6}
\end{center}
\end{figure*}

We now increase the Stokes number to 0.413 and the evolution of inertial particles is studied for different
spectral power laws as shown in Fig.~\ref{F5.6}. The reference case ($p=5/3$) for this particular value
of $St$ produces a 1D-H attractor. It corresponds to the second plot in row d of
Fig.~\ref{F5.6}.
We can notice that because of that relatively high inertia of the particles, the
attractor's shape is robust and does not change even when the power is increased or decreased in the range $]1.55-2[$.
This corresponds to the black (dark blue online) area around $Fr=0.85$ that
can be observed in Fig~\ref{F5.6}(a).

In agreement with the case $St=0.207$, we also observe that the dark gray (blue online) area shown around $Fr=0.85$ shifts towards the right as $p$ is increased (dark gray (blue online) spots for $0.8 \leqslant Fr < 0.57$). By comparing the rows (b) and
(d) in Fig.~\ref{F5.6}, the trace of an identical  attractor ($Fr=0.85$, $p=5/3$) can  be observed with higher power laws but at higher
gravity ($Fr=0.72$, $p=2.5$).

\begin{figure*}[htp]
\includegraphics [scale=0.37]{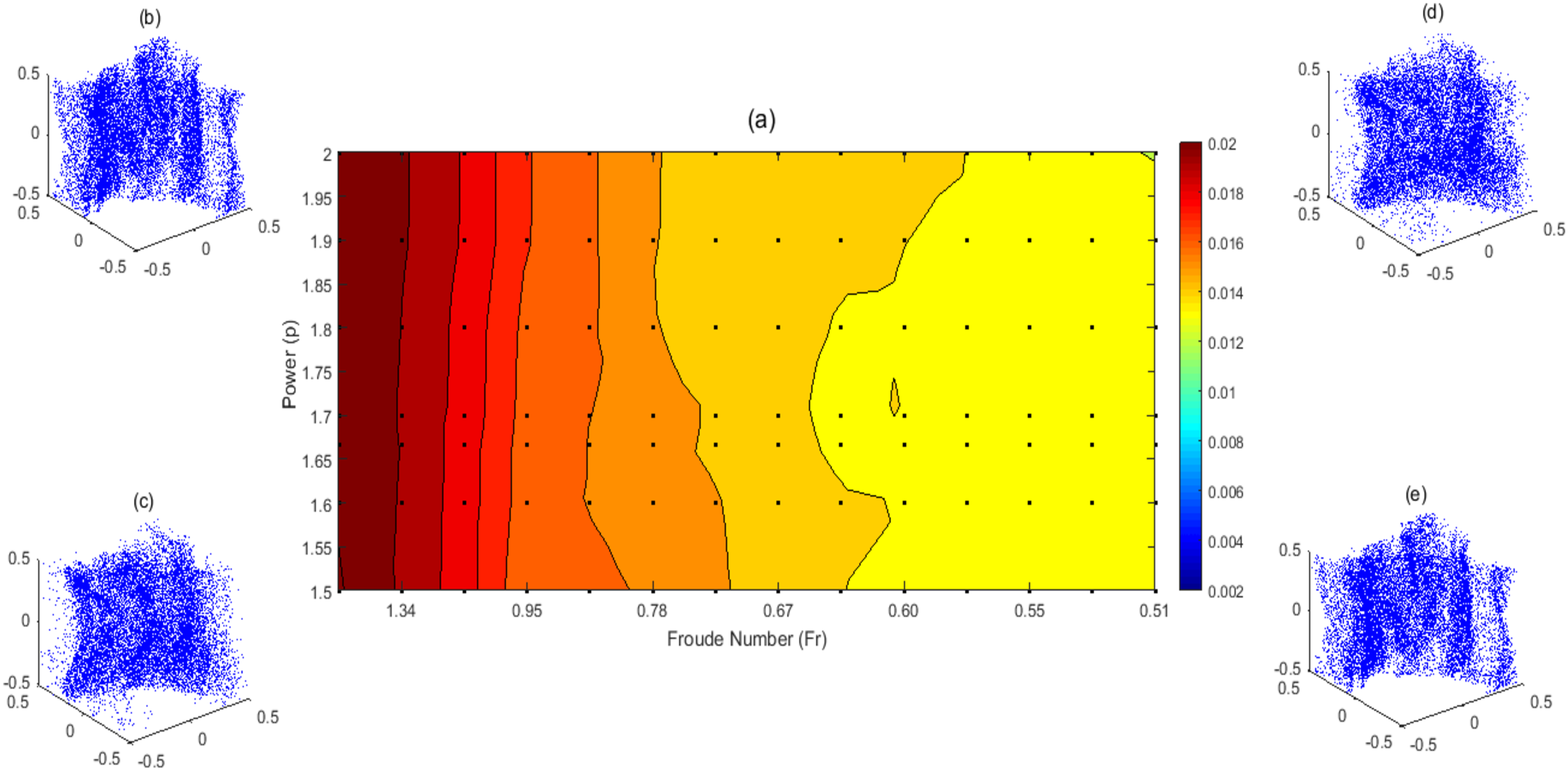}
\caption{\label{F5.7}
$St=1$:
(a)- Iso-contours of $\Delta$ as functions of
$(Fr,p)$. Typical particle clusters for different spectral power laws and
decreasing values of $Fr$:
(b) $Fr=0.67$, $p=1.67$,
(c) $Fr=0.95$, $p=1.67$,
(d) $Fr=0.95$, $p=2$ and
(e) $Fr=0.67$, $p=2$.
}
\end{figure*}

We also observe the 2D-L attractors at low $Fr$ which retain their shapes independently of $p$.
\\[2ex]
\noindent Finally, the attractor variations are examined for heavier particles
($St=1$, Fig.~\ref{F5.7}a). We note that there is no real deviation of the clustering
pattern for $p \neq 5/3$ as shown in Fig.~\ref{F5.7}(b)-(e).
This shows that gravity ($Fr$) is the main parameter
governing the clustering and the Eulerian structure becomes
less relevant as $St$ approaches unity.
\\[2ex]
The results are summarised in Table~\ref{Table-5.2}
where we add the case $St=0.124$ not shown here.
\begin{table}[htb]
\caption{\label{Table-5.2} Occurrences of attractors for given $St$ numbers varying $Fr$ and $p$.}
\begin{tabular}{|c|c|c|c|c|c|}
\hline \hline \multirow{2}{*}{Attractor}  & \multirow{2}{*}{\textbf{}} &
\multicolumn{4}{c|}{Stokes number $St$}                                          \\ \cline{3-6}
                                     &                                           & 0.124               &
0.207      & 0.413      & 1      \\ \hline \hline
\multirow{2}{*}{1D-H}      & $Fr$                      & 0.65-1.34           &
0.65-1.34          & 0.65-1.34           & \multirow{2}{*}{No} \\ \cline{2-5}
                                     & $p$                       & 1.5-2.5             &
1.5-2.0             & 1.5-2.5             &                     \\ \hline \hline
%
%
%
\multirow{2}{*}{1D-V}       & $Fr$                      & \multirow{2}{*}{No} &
0.5-0.65           & \multirow{2}{*}{No} & \multirow{2}{*}{No} \\ \cline{2-2} \cline{4-4}
                                     & $p$                       &                     &
1.5-2.             &                     &                     \\ \hline \hline
\multirow{2}{*}{2D-L}       & $Fr$                      & \multirow{2}{*}{No} &
\multirow{2}{*}{No} & 0.4-0.6           &   0.4-1.34                  \\ \cline{2-2} \cline{5-6}
                                     & $p$                       &                     &
& 1.5-2.5             &   1.5-2.5                  \\ \hline \hline
\multirow{2}{*}{1D-L} & $Fr$                      & \multirow{2}{*}{No} & \multirow{2}{*}{No}
            & \multirow{2}{*}{No} & \multirow{2}{*}{No} \\ \cline{2-2} 
                                     & $p$                       &                     & & &
\\ \hline \hline
\end{tabular}
\end{table}

\section{Conclusion}
\label{seconcl}

We quantify the variations in particle clustering in the presence of gravity with modified spectral power laws,
from very steep ($p \to 2.5$) to very flat ($p \to 1.4$)
energy distributions.

Though the existence and shape of a Lagrangian attractor depends on the three parameters $(Fr,St,p)$,
some general trends have been found.

The spectral law can have a significant effect
on the Lagrangian attractor shape but for some ranges of
Froude or Stokes numbers, in particular when a 2D Layer is achieved the spectral law has little or no effect on the
clustering.
1D or 2D attractors can only be observed when there is gravity
($Fr \neq \infty$).
In the absence of gravity ($Fr = \infty$) no 1D or 2D attractor is observed and the energy distribution ($p \in [1.4,2]$) has no effect
on this result.

However, the energy distribution can have an effect on 1D attractor shapes.
For instance, for the high values of the Froude number we studied ($Fr>1$), the orientation of the Lagrangian attractor
depends on the Stokes number $St$. But though the particles with low Stokes numbers, $St<0.2$, move towards a horizontal 1-D attractor with little effect of $p$, the variations in spectral power law do either modify or destroy the attractor structure
for larger Stokes numbers, $ 0.2< St \leqslant 0.5$.
This shows that when the gravity effect is small (large $Fr$),
the particles with larger $St$ are more sensitive
to a modification in the energy distribution.

As the gravity effect becomes more dominant ($0.5<Fr<0.95$),
the ranges of $St$ and $p$ which can lead to a one-dimensional attractor increases.

For low values of $Fr<0.5$,
no attractor develops
in the horizontal direction,
the large gravity effect is to stretch the attractors in the vertical direction and force the particles to
move in clear vertical patterns. As a result, we observed some 2D-L, 1D-V
or the complex 1D-L attractors.
This complex 1D-L attractors was
only observed for $p > 5/3$.

%

\section{Acknowledgement}

This work was supported by EPSRC grant EP/L000261/1

%


\end{document}